\documentclass[prb,aps,twocolumn,floats,showpacs]{revtex4}
\usepackage{graphicx}
\usepackage{amsmath}

\newcommand{\be}{\begin{equation}}
\newcommand{\ee}{\end{equation}}
\newcommand{\bea}{\begin{eqnarray}}
\newcommand{\eea}{\end{eqnarray}}

\newcommand{\lb}{\left[}
\newcommand{\rb}{\right]}
\newcommand{\lp}{\left(}
\newcommand{\rp}{\right)}
\newcommand{\lf}{\left\{}
\newcommand{\rf}{\right\}}

\newcommand{\E}{{\cal E}}

\newcommand{\Ry}{{\rm Ry}\,}
\newcommand{\EF}{{\epsilon_F}}
\newcommand{\ts}[1]{\textstyle{#1}}

\begin{document}
\title{Temperature-dependent Drude transport in a two-dimensional electron gas}
\author{D. S. Novikov}
\affiliation{Department of Physics, Yale University, New Haven, Connecticut 06520}
\date{February 14, 2009}


\begin{abstract}

We consider transport of dilute two-dimensional electrons,
with temperature between Fermi and Debye temperatures. 
In this regime, electrons form a non-degenerate plasma 
with mobility limited by potential disorder. Different kinds of 
impurities contribute unique signatures to the resulting 
temperature-dependent Drude conductivity, via energy-dependent scattering.
This opens up a way to characterize sample disorder composition.
In particular, neutral impurities cause a slow decrease of conductivity 
with temperature, whereas charged impurities result in conductivity growing as 
a square root of temperature. This observation serves as a precaution for 
literally interpreting metallic or insulating conductivity dependence, 
as both can be found in a classical metallic system.

\end{abstract}
\pacs{
      72.10.-d	
      73.40.-c	
      71.30.+h 	
      81.05.Uw	
}

\maketitle

\section{Introduction}

Electron transport is conventionally understood 
within the Fermi liquid theory framework.\cite{AA}
In a Fermi liquid (FL), screening of the offset charge 
is very efficient due to large density of states. As a result, 
effective disorder potential for quasiparticles is always short-ranged.
The latter leads to temperature-independent Drude
conductivity, while interaction effects provide corrections to 
the Drude transport in the powers of small parameter $T/\EF$, where 
$T$ is temperature ($k_B=1$), and  $\EF$ is the Fermi energy.
These corrections originate 
due to scattering off the Friedel oscillations,\cite{AA,ZNA}
and due to temperature dependence of the random-phase approximation (RPA) 
screening.\cite{GD}

\begin{figure}[b]
\includegraphics[width=3.5in]{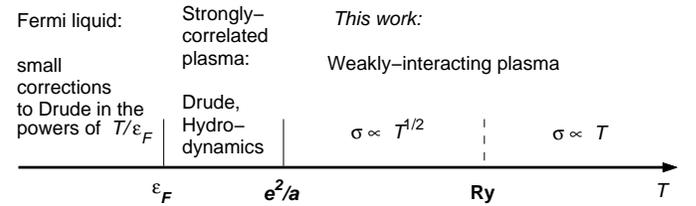}
\caption[]{
2D transport in the presence of charged disorder for $r_s=(e^2/a)/\EF>1$. 
FL calculations \cite{AA,ZNA,GD} provide small corrections to 
the Drude transport at $T\ll \EF$. For non-degenerate carriers,
the perturbative result  $\sigma\propto T$ \cite{DSH} is valid for 
$T\gg \Ry, e^2/a$, while for $e^2/a \ll T \ll \Ry$ we show that 
$\sigma \propto T^{1/2}$. For $\EF \lesssim T \lesssim e^2/a$ the system forms
a strongly-correlated plasma where both Drude \cite{Dykman}
and hydrodynamic \cite{Spivak} effects can be relevant.
For $e^2/a < \EF$, i.e. $r_s<1$,  
weakly-interacting Fermi gas crosses over to weakly-interacting classical 
plasma at $T\sim \EF$; the result $\sigma\propto T^{1/2}$ then holds
for $\EF \ll T \ll \Ry$.
}
\label{fig:cond-regimes}
\end{figure}

What happens when the carrier density $n$ is reduced so much that 
the system becomes non-degenerate, $T> \EF$? In this case the Fermi energy 
is irrelevant, and the system is essentially a classical plasma 
(Fig.~\ref{fig:cond-regimes}). 
Such a situation may occur in actively studied clean dilute heterostructures
\cite{Hanein,Mills,Lilly,Noh,Jian,Manfra,Ho} where $n\lesssim 10^{10}\,$cm$^{-2}$, 
with record densities down to $7\times 10^8\,$cm$^{-2}$ (Ref.~\onlinecite{Jian}). 
These densities correspond to $\EF\sim 10-100\,\mu$V$\sim 0.1-1\,$K.
Transport in such a system will depend on the strength $e^2/a$ 
of electron interactions relative to temperature (here 
$a=1/\sqrt{\pi n}$ is the Wigner-Seitz radius, 
and the dielectric constant $\kappa$ is included into the definition 
of charge, $e^2 \to e^2/\kappa$, for brevity).
For strong interactions, $e^2/a > T$, we have a strongly-correlated
semiclassical electron ``liquid'' 
whose collective modes are likely to affect transport.\cite{Dykman,Spivak}

Here we consider the two-dimensional (2D) 
transport in the opposite, weakly-interacting regime, 
\be \label{hi-T}
\EF, \ e^2/a \ll  T \ll \Theta_D \,. 
\ee
We assume that temperature is high enough so that carriers form 
a classical weakly-interacting plasma, 
yet is well below the Debye temperature $\Theta_D$ so that 
the phonon contribution to transport can be either neglected or subtracted
in a controlled way.
Transport is then dominated by the practically {\it unscreened} potential disorder.
Such a situation can become relevant in the cleanest  
heterostructures (e.g. Ref.~\onlinecite{Jian}), where, 
for the lowest densities, the interaction energy
$e^2/a \approx 5\,$K is one order of magnitude below the Debye temperature. 
With increasing sample quality, the carrier density decreases
and the applicability range (\ref{hi-T}) widens. Another system where
the present approach may be applicable is graphene with the 
substrate-induced gap, as described towards the end of the paper.

We show that in the regime (\ref{hi-T}), 
the Drude conductivity $\sigma(T)$ becomes strongly 
temperature-dependent. Its temperature dependence originates from 
the energy-dependent impurity scattering cross-section. Remarkably, 
different kinds of potential impurities (e.g. charged, neutral) 
can now be distinguished by qualitatively different energy dependence of scattering, 
yielding unique signatures in the resulting $\sigma(T)$. 
These signatures could be used to characterize notoriously unknown potential 
profile for high-quality 2D samples.

In particular, for the important example of {\it charged impurities} within 
a 2D layer, we show that the conductivity grows as $\sigma \propto \sqrt{T}$
as long as temperature is below a few Rydberg of the host material 
($\Ry = me^4/2\hbar^2$ where $m$ is the effective carrier mass), 
crossing over to $\sigma \propto T$ for $T\gg \Ry$ (Fig.~\ref{fig:cond-regimes}).
The latter linear $T$-dependence\cite{DSH} is thereby practically inobservable 
for two-dimensional electron gases (2DEGs) 
in GaAs heterostructures, since the phonon contribution 
dominates above $\Theta_D \approx \Ry \approx 60\,$K.
Hence, the single-particle explanation of Das Sarma and Hwang\cite{DSH}
of the observed\cite{Hanein,Mills,Lilly,Noh} 
conductivity increase with temperature does not apply.
For the other practical example, the strong {\it neutral impurities}, 
the conductivity is shown to {\rm decrease} with temperature (as described below).

As a result, the superficial distinction between a ``metal'' ($d\sigma/dT<0$) 
and an ``insulator'' ($d\sigma/dT>0$) based simply on the {\it sign} of the 
derivative $d\sigma/dT$, does not hold -- indeed, both behaviors are possible 
in a classical 2D metal (\ref{hi-T}). 
Of course, a true insulator is characterized by localized states as $T\to 0$,
leading to activated conductivity dependence. Such a low-$T$ analysis is beyond 
the scope of this work which considers only sufficiently high temperatures
above the onset of localization.

In what follows, we first obtain the general result (\ref{cond-T}) 
for the $T$-dependent Drude conductivity in the regime (\ref{hi-T}) 
in terms of the energy-dependent
impurity transport cross-section $\Lambda_{\rm tr}(\epsilon)$, 
then we discuss the resulting $\sigma(T)$ for different kinds of potential 
disorder, and, finally, remark on the systems where 
one can practically observe the temperature-dependent Drude conductivity.

\section{The Drude transport}

The kinetic equation in the presence of an external in-plane field $\vec{\E}$
\be \label{KE-e-dep}
e\vec{\cal E}{\bf v} \partial_\epsilon f_0 = 
- \tau_{\epsilon}^{-1} \delta f  \,, \quad \epsilon=mv^2/2 
\ee
is written in terms of the momentum relaxation rate
\be \label{tau-lambda}
\tau_\epsilon^{-1} = n_i \Lambda_{\rm tr}(\epsilon)v(\epsilon) \,, 
\quad \Lambda_{\rm tr}= \oint\! d\theta 
{d\Lambda\over d\theta}  
(1-\cos\theta) \,.
\ee
Here $d\Lambda/d\theta$ is the differential scattering cross-section and 
$n_i$ is the area density of impurities.
The particular energy dependence of the scattering rate $\tau_\epsilon^{-1}$
stems from that of the transport cross-section $\Lambda_{\rm tr}$.
The isotropic dc conductivity follows\cite{AFS}:
\be \label{drude-e}
\sigma = {ne^2\bar\tau\over m} \,, \quad
\bar\tau = 
\frac{\int \! d\epsilon\, \epsilon \tau_\epsilon 
(-\partial_\epsilon f_0)}
{\int \! d\epsilon \, \epsilon (-\partial_\epsilon  f_0)} \,,
\ee
where we assumed energy-independence of the 2D density of states 
in the case of the parabolic band.
In the classical regime (\ref{hi-T}), quantum interference effects
are irrelevant due to strong dephasing. As long 
as $T\gg e^2/a$, one can also neglect electron-electron interactions, such that
the equilibrium velocity distribution is Maxwellian: 
\be \label{Maxwell}
f_0\big(\epsilon(v)\big) 
= e^{(\mu-\epsilon)/T}\,, \quad e^{\mu/T} = {\EF/ T} \ll 1\,.
\ee
Eqs.~(\ref{drude-e}) and (\ref{Maxwell}) yield
\be \label{cond-T}
\sigma(T) = \sigma_0 \int_0^\infty \! \xi d\xi \, e^{-\xi} 
\left. {\lambda(\epsilon)\over \Lambda_{\rm tr}(\epsilon)}\right|_{\epsilon=\xi T} , 
\quad \sigma_0 \equiv  {e^2\over h}  {n\over n_i} \,,
\ee
where energy-dependent wavelength 
$\lambda(\epsilon)=2\pi/k(\epsilon)$, $\hbar k = mv(\epsilon)$.
In other words, the temperature dependence of the Drude conductivity 
is determined by the energy dependence 
of the transport cross-section in the units of 
wavelength. For simple estimates, Eq.~(\ref{cond-T}) gives 
\be \label{cond-estimate}
\sigma(T) \simeq \sigma_0 
{\lambda_T \over \Lambda_{\rm tr}|_{\epsilon = T}} \,,
\ee
where $\lambda_T = 2\pi\hbar/\sqrt{2mT}$ is the temperature wavelength.

When multiple kinds of impurities are present, 
the scattering rates add up according to the Matthiessen rule.
Thus the total transport cross-section entering Eq.~(\ref{cond-T})
\be \label{matthiessen}
\Lambda_{\rm tr}(\epsilon)  =
c_1 \Lambda_{\rm tr}^{(1)}(\epsilon) + c_2 \Lambda_{\rm tr}^{(2)}(\epsilon) + ...
\ee
where $c_j = n_{i}^{(j)}/n_i$ is the fraction of impurities of the sort $j$,
and $n_i = \sum_j n_{i}^{(j)}$ is the total impurity concentration.

\begin{figure}[t]
\includegraphics[width=3.5in]{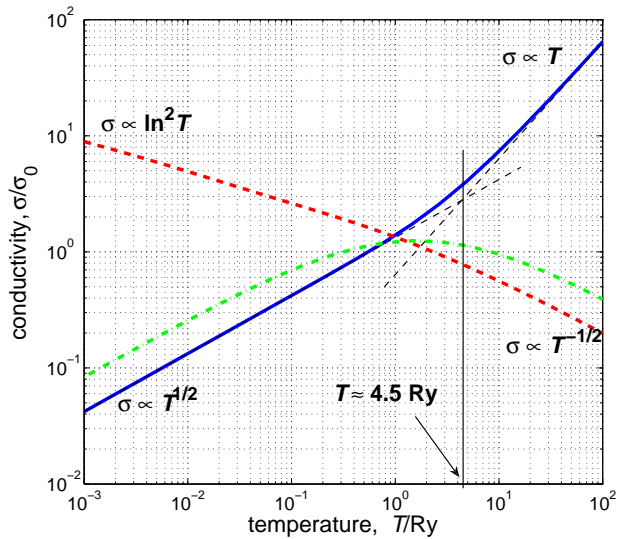}
\caption[]{Temperature-dependent Drude conductivity (\ref{cond-T})
in the units of $\sigma_0 = (e^2/h) n/n_i$.
Solid blue line: charged impurities, Eq.~(\ref{cond-ruth-2d}), together with 
its asymptotic limits (thin dashed lines).
Switching from $\sigma \propto T^{1/2}$ to $\sigma\propto T$ occurs at 
$T\approx 4.5\,{\rm Ry}$. Dashed red line: neutral impurities 
(impermeable disks with radius $a=a_B$).
Dash-dotted green line: 50\% charged and 50\% neutral impurities,
with the same total $n_i$.
}
\label{fig:T_dep_cond}
\end{figure}

\section{Charged impurities}

For the $e^2/r$ potential, the exact 2D differential
cross section has been found in the seminal 1967 work of Stern and 
Howard \cite{Stern-Howard}:
\be \label{ruth-2d}
{d\Lambda^{\rm c} \over d\theta} = 
{\alpha \tanh \pi\alpha \over 2k \sin^2(\theta/2)} \,, \quad
\alpha(v) \equiv {e^2\over \hbar v} \,.
\ee
Here $\theta$ is the scattering angle, and the momentum transfer 
$q = 2\hbar k \sin \frac\theta2$.
The result (\ref{ruth-2d}) has two distinct limits.
For small energies, $\epsilon \ll \Ry$, the parameter $\pi\alpha \gg 1$, and
the cross section is classical (indeed, it is $\hbar$-independent when 
$\tanh \pi\alpha \equiv 1$).
Conversely, for high energies ($\epsilon\gg \Ry$), the cross-section
(\ref{ruth-2d}) with $\tanh \pi\alpha \approx \pi\alpha$ 
coincides with the Born approximation. 
Such a classical-to-quantum crossover is specific 
to 2D, whereas in 3D the Rutherford cross-section coincides both with 
the classical result and with the Born approximation.\cite{Landau3}

The corresponding 2D transport cross-section (\ref{tau-lambda}) reads
\be \label{lambda-t-2d}
\Lambda^{\rm c}_{\rm tr} = (2\pi\alpha/k) \tanh \pi\alpha \,.
\ee
Notably, it is finite, with all scattering angles
contributing roughly equally. This should be contrasted with
the well-known logarithically divergent transport cross-section 
for 3D Coulomb plasma \cite{Landau10} (``Landau logarithm''),
dominated by forward scattering processes.

The conductivity then follows from Eqs.~(\ref{cond-T}) and (\ref{lambda-t-2d}):
\be \label{cond-ruth-2d}
\sigma^{\rm c}(T) =
\sigma_0 \int_0^\infty\, {\xi d\xi e^{-\xi} \over \alpha \tanh \pi\alpha}\,,
\quad \alpha^2 = {\Ry\over \xi T}\,.
\ee
The conductivity (\ref{cond-ruth-2d})
grows with $T$ (Fig.~\ref{fig:T_dep_cond}) since impurity scattering 
(\ref{ruth-2d}) weakens for faster moving carriers.

The asymptotic behavior of (\ref{cond-ruth-2d}) is
$\sigma^{\rm c} \simeq \ts{\frac34}\sigma_0\sqrt{\pi T/\Ry}$ at $T\ll \Ry$, 
and $\sigma^{\rm c} \simeq \frac2\pi\sigma_0 T/\Ry$ at $T\gg \Ry$.
Practically, the switching between the two limits occurs when 
$T\approx 4.5\,\Ry$ (Fig.~\ref{fig:T_dep_cond}, the two asymptotes cross).
For $T\gtrsim 5\,\Ry$ we agree with Ref.~\onlinecite{DSH} 
where the Born approximation 
$\tau_\epsilon^{-1} \approx {n_i\pi^2 e^4/\hbar \epsilon}$ was utilized
in Eq.~(\ref{tau-lambda}) [corresponding to the Fermi Golden Rule]. 
The limit $\sigma \propto \sqrt{T}$ is novel and  
relevant in the parameter range (\ref{hi-T}).

\section{Neutral scatterers}

For any axially symmetric scatterer, the transport cross-section is given by
\cite{Stern-Howard}
\be \label{lambda-tr-delta}
\Lambda_{\rm tr} ={2\over k}\sum_{m=-\infty}^\infty \sin^2(\delta_{m+1}-\delta_m)\,, 
\ee
where $\delta_m$ is the scattering phase shift in the channel with orbital
momentum $m$.

Consider an example of {\it strong neutral 2D scatterers} within the electron layer.
Physically, they can originate from  
interface roughness or neutral atomic defects in a heterostructure.
For sufficiently small $\EF$ and $T$, the carrier's energy may become much 
smaller than the potential barrier which such a potential creates. 
It is then reasonable to model the scattering potential as being infinitely
large within a disk of radius $a$, and zero outside.
In this case, the scattering phase shifts $\tan \delta_m = J_m(ka)/Y_m(ka)$
are given in terms of the Bessel functions of the first and second kind,
leading to 
\be \label{lambda-disk}
\Lambda_{\rm tr}^{\rm n} 
\simeq \lf
\begin{matrix} 
{8a/ 3}\,, &  ka\gg 1 \,; \\
{ \pi^2/ k\over {\pi^2/ 4} + \ln^2 \lb {2/ (\gamma ka)}\rb}\,,
&  ka\ll1\,.
\end{matrix}
\right.
\ee
Here $\ln\gamma = 0.577...$ is Euler's constant. 
Note the anomalously efficient scattering at wavelengths  
$\lambda = 2\pi/k\gg a$ exceeding the impurity size: 
$\Lambda_{\rm tr}^{\rm n} \sim \lambda/\ln^2(\lambda/a) \gg a$, i.e. the scattering
cross-section is determined by the carrier wavelength rather than by the 
impurity size, thereby greatly exceeding the ``geometric'' limit. This is a known 
universal signature of low-energy 2D scattering.\cite{Landau3}

The estimate (\ref{cond-estimate}) yields
\be \label{cond-T-2d}
\sigma^{\rm n}(T) \sim \sigma_0 \times {\rm min } 
\lf \ln^2{\epsilon_a\over T}, \ \lp{\epsilon_a\over T}\rp^{\frac12}\rf , \quad 
\epsilon_{a} \sim {\hbar^2\over ma^2}. 
\ee
The exact conductivity $\sigma^{\rm n}(T)$ for strong neutral scatters
calculated numerically using Eqs.~(\ref{lambda-tr-delta}) and (\ref{cond-T}), is 
shown in Fig.~\ref{fig:T_dep_cond}. Its asymptotic behavior for small and 
large $T$ agrees with the qualitative estimate (\ref{cond-T-2d}).
In order to compare with the Coulomb scattering, we took the disk radius
$a=a_B$ to be equal to the Bohr radius $a_B=\hbar^2/me^2$, 
such that $\epsilon_a \sim\Ry$; $a_B\sim 10\,$nm for GaAs.

We also note that {\it weak short-range scatterers} yield 
temperature-independent conductivity. Indeed, the differential cross-section
$d\Lambda/d\theta = |f(\theta)|^2$ in the Born approximation 
$f^{\rm Born}(\theta) = -m\tilde U(q)/\hbar^2\sqrt{2\pi k}$ 
(Ref.~\onlinecite{Landau3}) yields 
$\Lambda_{\rm tr}^{\rm Born} \propto \lambda$ for $q$-independent 
formfactor $\tilde U(q)$ corresponding to a short-ranged potential $U(r)$.
Eq.~(\ref{cond-T}) then results in $\sigma=2\pi \sigma_0 (\hbar^2/m\tilde U)^2$=const.

\section{Disorder spectroscopy}

In realistic clean low-density samples multiple kinds of  
disorder, e.g. Coulomb impurities and neutral scatterers, are present. 
The conductivity (\ref{cond-T}) and (\ref{matthiessen})
can then display a fairly complex sample-specific dependence on 
temperature, governed by relative contributions of different kinds of 
scatterers. Fig.~\ref{fig:T_dep_cond} shows an example with 
$c^{\rm c} = c^{\rm n} = 0.5$.

Qualitatively different $T$-dependences
(\ref{cond-ruth-2d}) and (\ref{cond-T-2d}) present a natural way to characterize  
disorder in clean 2D samples. For that one needs to operate at very low carrier
densities $n \lesssim 10^9$\,cm$^{-2}$, when a temperature window (\ref{hi-T}) 
opens up. Fitting the conductivity (with the phonon contribution subtracted) 
to the result 
(\ref{cond-T}) and (\ref{matthiessen}) will yield the disorder composition
$\{n_i^{(j)} \}$.
This way, the $T$-dependent transport can serve as the disorder spectroscopy. 
The connection with spectroscopy is not accidental:  
Formally, the conductivity is proportional to Laplace transform 
$\int_0^\infty \! d\epsilon\, e^{-\beta \epsilon} \varphi(\epsilon)$
of the quantity $\varphi(\epsilon) = \sqrt{\epsilon}/\Lambda_{\rm tr}(\epsilon)$, 
where $\beta = 1/T$.

\section{Charged disorder in G\lowercase{a}A\lowercase{s} heterostructures}

The result (\ref{cond-ruth-2d}) based on the exact 
cross-section (\ref{ruth-2d}), predicts a novel
$\sigma \propto \sqrt{T}$ conductivity dependence, characteristic 
of the classical limit of scattering (\ref{ruth-2d}). 
The latter can be relevant for transport in 
clean dilute heterostructures.\cite{Hanein,Mills,Lilly,Noh,Jian,Manfra,Ho}
So far, the observed conductivity grows approximately linearly with temperature
around $T\sim 1\,$K.\cite{Hanein,Mills,Lilly,Noh} 
From the present analysis, the single-particle
explanation \cite{DSH} for this observation based on the Born scattering 
cannot hold for GaAs, since, according to Fig.~\ref{fig:T_dep_cond}, 
the crossover to the Born regime would occur at $T \approx 300\,$K 
which is practically inaccessible due to the dominant 
phonon scattering.\cite{Gao'05,Ridley-review} 
The apparent discrepancy between the present  
single-particle theory yielding $\sigma\propto \sqrt{T}$,  
and the experiments \cite{Hanein,Mills,Lilly,Noh}
strongly indicates the predominance of collective effects in transport.
This is not surprising, since typical Coulomb energy $e^2/a\simeq 20\,$K
[corresponding to $n=1\times 10^{10}\,$cm$^{-2}$], while the measurements 
were done for at least order-of-magnitude lower temperatures, in which case using
the Maxwell distribution (\ref{Maxwell}) in Eq.~(\ref{drude-e}) is 
unjustified from the outset. For the lower densities, $n\lesssim 10^9\,$cm$^{-2}$,
the present approach may apply, as long as the phonon contribution 
is controllably subtracted in the range (\ref{hi-T}).

Can the linear (RPA) screening affect the temperature dependence (\ref{cond-ruth-2d}),
and in particular, the crossover temperature $T\approx 5\,\Ry$?
Below we argue that screening will only {\it weaken} the dependence $\sigma(T)$,
and cannot lead to $\sigma(T)\propto T$ at low temperature.

Physically, screening changes the { shape} of the 
impurity potential in the following way: 
It fully preserves the strength of the $e^2/r$ potential for distances 
$r\lesssim a_s$ shorter than the screening length, and cuts off the $1/r$ behavior 
for $r\gtrsim a_s$, where $a_s = \frac{T}{2\pi e^2 n} = \frac{a}2  {T\over e^2/a}$
[in the Fourier space, $2\pi e^2/k \to 2\pi e^2/(k+a_s^{-1})$].
The linear (RPA) screening is a mean-field effect,
valid when the density fluctuations within the screening volume $a_s^2$
are small, fulfilled under the condition $n a_s^2 \gg 1$ equivalent to 
$T\gg {e^2/a}$, compatible with the limit (\ref{hi-T}).
This has the following consequences: 
(i) for $\EF < T < {e^2/ a}$ relying on the RPA screening 
is unjustified. The single-particle transport calculation based on the 
Maxwell distribution (\ref{Maxwell}) is also unjustified.
Thus the approach\cite{DSH} of Das Sarma and Hwang does not apply to the 
experiments\cite{Hanein,Mills,Lilly,Noh} even 
if the authors were to use the correct scattering cross-section.
(ii) For $T \gg {e^2/a}$, screening becomes asymptotically 
{\it irrelevant} for the Drude transport.
Indeed, consider the region  $r < a_s$ where the electron ``feels'' 
the unscreened impurity potential. Upon entering this region, 
its typical kinetic energy greatly exceeds the Coulomb field, 
$T\gg e^2/a_s$. Thus the scattering phase shifts
yielding the cross-section (\ref{ruth-2d}) have parametrically large 
room to accumulate between $e^2/T \ll r \ll a_s$, leading to its nonperturbative
limit.
Moreover, the residual screening (truncation of the potential for 
$r\gtrsim a_s$) would
further weaken the $\sigma(T)$ dependence, since, according to the above 
calculation [cf. Eqs.~(\ref{cond-T}) and (\ref{cond-T-2d})], 
the conductivity due to short-range disorder decreases with temperature.
Thus the initial $\sigma\propto \sqrt{T}$ dependence would only weaken 
when the residual screening is taken into account.

As a result, the explanation \cite{DSH} suggested for the apparent linear growth 
of the conductivity with temperature,\cite{Hanein,Mills,Lilly,Noh}
does not apply; the observed linear (and, generally, power-law\cite{Jian}) 
$T$-dependence of the low-temperature conductivity remains an exciting 
unresolved problem.

\section{Graphene with charged disorder}

The nonrelativistic scattering considered above can be 
applied to graphene samples where Dirac mass $m=\Delta/v_F^2$ can originate 
e.g. from symmetry-breaking between sublattices, such that gap 
$\Delta\sim 10-100\,$meV.\cite{khomyakov,lanzara-gap}
Half-filled band corresponds to chemical potential $\mu=-\Delta$ counted 
from the bottom of the ``parabolic'' band.
The graphene electron system is nonrelativistic and nondegenerate 
as long as $T\ll\Delta$, since $\EF = T e^{-\Delta/T}\ll T$ [Eq.~(\ref{Maxwell})].
When electron interactions (controlled by dielectric environment) are weak,
$\alpha_g=e^2/\hbar v_F \ll 1$ where $v_F\simeq 10^6\,$m/s, the 
effective Rydberg ${\rm Ry}_g = \alpha_g^2\Delta/2 \ll \Delta$. 
Hence, cf. Fig~\ref{fig:cond-regimes}, 
the conductivity $\sigma \propto T^{1/2}$ for $T \ll {\rm Ry}_g$ and 
$\sigma \propto T$ for ${\rm Ry}_g \ll  T \ll \Delta$. 
For strong interactions, $\alpha_g \sim 1$, 
${\rm Ry}_g \sim \Delta$ and the regime $\sigma\propto T$ never plays out. 
For $T\gg \Delta$ the system becomes relativistic, the 
cross-section scales as the wavelength,\cite{graphene-scatt} 
and the $T$-dependence of the Drude conductivity $\sigma \propto T^2$ 
comes solely from that of carrier density $n\propto T^2$, 
Ref.~\onlinecite{graphene-asym}.

\section{Summary}

To conclude, we considered temperature-dependent Drude transport
in non-degenerate 2D electron systems.
The Drude conductivity due to charged disorder 
behaves classically, $\sigma\propto T^{1/2}$ for  
temperatures below a few Rydberg, while neutral disorder results in decreasing 
$\sigma(T)$. These signatures can be utilized in determining disorder content 
of clean 2D samples in the limit (\ref{hi-T}). 
The decrease of the conductivity while reducing temperature 
does not necessarily signify a transition to an insulating state.

\acknowledgments
This work has benefited from discussions with M. Dykman and L. Glazman.
Research was supported by NSF grants No. DMR-0749220 and No. DMR-0754613.


\end{document}